\documentclass[preprint,superscriptaddress,aip,onecolumn,12pt]{revtex4-1}
\usepackage{graphicx}
\usepackage{dcolumn}
\usepackage{bm}
\usepackage{amsmath}
\usepackage[english]{babel} 
\usepackage{hyperref}

\begin{document}

\title{Design for ultrahigh-$Q$ position-controlled nanocavities of single semiconductor nanowires in 2D photonic crystals}
\author{Muhammad Danang Birowosuto}
\email{birowosuto.danang@lab.ntt.co.jp}
\affiliation{NTT Basic Research Laboratories, NTT Corporation, 3-1 Morinosato Wakamiya, Atsugi, Kanagawa 243-0198, Japan}
\affiliation{NTT Nanophotonics Center, NTT Corporation, 3-1 Morinosato Wakamiya, Atsugi, Kanagawa 243-0198, Japan}
\author{Atsushi Yokoo}
\affiliation{NTT Basic Research Laboratories, NTT Corporation, 3-1 Morinosato Wakamiya, Atsugi, Kanagawa 243-0198, Japan}
\affiliation{NTT Nanophotonics Center, NTT Corporation, 3-1 Morinosato Wakamiya, Atsugi, Kanagawa 243-0198, Japan}
\author{Hideaki Taniyama}
\affiliation{NTT Basic Research Laboratories, NTT Corporation, 3-1 Morinosato Wakamiya, Atsugi, Kanagawa 243-0198, Japan}
\affiliation{NTT Nanophotonics Center, NTT Corporation, 3-1 Morinosato Wakamiya, Atsugi, Kanagawa 243-0198, Japan}
\author{Eichii Kuramochi}
\affiliation{NTT Basic Research Laboratories, NTT Corporation, 3-1 Morinosato Wakamiya, Atsugi, Kanagawa 243-0198, Japan}
\affiliation{NTT Nanophotonics Center, NTT Corporation, 3-1 Morinosato Wakamiya, Atsugi, Kanagawa 243-0198, Japan}
\author{Masato Takiguchi}
\affiliation{NTT Basic Research Laboratories, NTT Corporation, 3-1 Morinosato Wakamiya, Atsugi, Kanagawa 243-0198, Japan}
\affiliation{NTT Nanophotonics Center, NTT Corporation, 3-1 Morinosato Wakamiya, Atsugi, Kanagawa 243-0198, Japan}
\author{Masaya Notomi}
\affiliation{NTT Basic Research Laboratories, NTT Corporation, 3-1 Morinosato Wakamiya, Atsugi, Kanagawa 243-0198, Japan}
\affiliation{NTT Nanophotonics Center, NTT Corporation, 3-1 Morinosato Wakamiya, Atsugi, Kanagawa 243-0198, Japan}
\date{\today}% It is always \today, today,

\begin{abstract}
Using Finite-Difference Time-Domain (FDTD) simulation, we show that ultrahigh-$Q$ nanocavities can be obtained through the manipulation of a single semiconductor nanowire (NW) inside a slot in a line defect of a two-dimensional (2D) photonic crystal. By controlling the design and its lattice parameters of the photonic crystal, we have achieved a quality factor $Q$ larger than $10^{6}$ and a mode volume $V_{c}$ smaller than 0.11 $\mu$m$^{3}$ (1.25 of a cubic wavelength in the NW) for a cavity peak in the telecommunication band. This design is useful for realizing a position-controlled cavity in a photonic crystal. Here we also discuss the small dependence of the $Q$-factor, the $V_{c}$, and the cavity peak in relation to the position of the NW inside the slot and the potential application to the cavity quantum electrodynamics (QED) using the embedded-emitter NW.
\end{abstract}

\maketitle

\section{Introduction}

In recent years, many new types of optical cavities with a high quality factor ($Q$) and a small mode volume ($V_{c}$) have been developed and fabricated \cite{Vahala2003}. From among these designs, photonic crystal slab nanocavities have attracted a lot of attention since they offer a very high $Q/V_{c}$ and the capacity for integration with other devices through nanowaveguides \cite{Notomi2010}. These nanocavities may also exhibit an ultrahigh-$Q (>10^{6})$ and a small $V_{c} (\sim(\lambda/n)^{3})$ \cite{BSSong2005,Kuramochi2006,Tanabe2007}, and this has led to the proposal and demonstration of ultralow power devices including power switches \cite{Shinya2008,Nozaki2010}, ultralow threshold lasers \cite{Matsuo2010,Ellis2011}, and all-optical random access memory (RAM) \cite{Nozaki2012}. Thus, the high-$Q$ nanocavities may enable the dense integration of ultralow power photonic devices in a single chip \cite{Notomi2004}.

Most ultrahigh-$Q$ nanocavities are fabricated by a combination of high-resolution electron beam lithography and dry etching and the nanocavities must be incorporated in the photonic crystal design prior the fabrication process. But recently, the high-$Q$ nanocavities have been obtained through a post modification of photonic crystal waveguides by employing scanning probe lithography \cite{Yokoo2011}. Some high-Q ($\lesssim 10^{4}$) nanocavities can also be created by filling the holes in the photonic crystals via water or polymer micro-infiltration \cite{Intonti2006,Intonti2009} but the feasibility of realizing a liquid-filled cavity is relatively low. Also, although these methods provide a rewritable cavity, there is no method with the ability to control the position of the cavity. Such a movable cavity would be useful in on-demand position-controlled devices in photonic integrated circuits.

A single semiconductor nanowire (NW) in the line defect of a two-dimensional (2D) photonic crystal may offer the possibility of realizing a position-controlled cavity since a single NW may modify the refractive index at any position in the line defect. The coupling of a single semiconductor NW embedded in a one-dimensional (1D) photonic crystal is intended solely to reduce the cavity losses of the NW caused by the reduction of the evanescent field outside the NW body \cite{Barrelet2006,Zhang2008}. Furthermore, the coupling of an NW with a 2D photonic crystal waveguide is designed solely for efficient photon delivery \cite{Grillet2007,Park2008}. For coupling between a cavity and an emitter, NWs with an embedded single emitter (e. g. quantum dots (QDs) \cite{Dorenbos2010,Claudon2010,Dalacu2011,Bleuse2011,Reimer2012,Bulgarini2012} and diamond color centers \cite{Babinec2010}) already exhibit excellent performances as single photon sources. Therefore, in addition to the position-controlled cavity, the incorporation of a single embedded emitter in a single NW may also provide a site-controlled emitter for a light matter interaction. This position-controlled cavity and emitter could be interesting as regards the revolution taking place in the fields of quantum computation \cite{Knill2001}, quantum cryptography \cite{Gisin2002}, and quantum optics \cite{Strauf2010}.

In this work, we propose high-$Q$ nanocavities based on a single semiconductor NW inside the air slot of a line defect in a 2D photonic crystal, and we demonstrate nanocavities with $Q \simeq 2 \times 10^{6}$ and $V_{c} < 0.11 \mu$m$^{3}$ (1.25 of a cubic wavelength in the NW) for the cavity peak in the telecommunication band. Our approach shows the possibility of engineering the position of the NW through NW manipulations. We use a numerical modeling with the three-dimensional Finite-Difference Time-Domain (3D FDTD) method to show that our platform still exhibits a relatively large $Q$ and small $V_{c}$ despite abrupt changes in the NW size, slot depth, slot width, and NW position. We also estimate that the embedded QD NWs are well suited for cavity quantum electrodynamics (QED) experiments by considering the light matter interaction with the single QD at the center of the NW.

\section{Cavity Design}

\begin{figure}[htbp]
\centering\includegraphics[width=13cm]{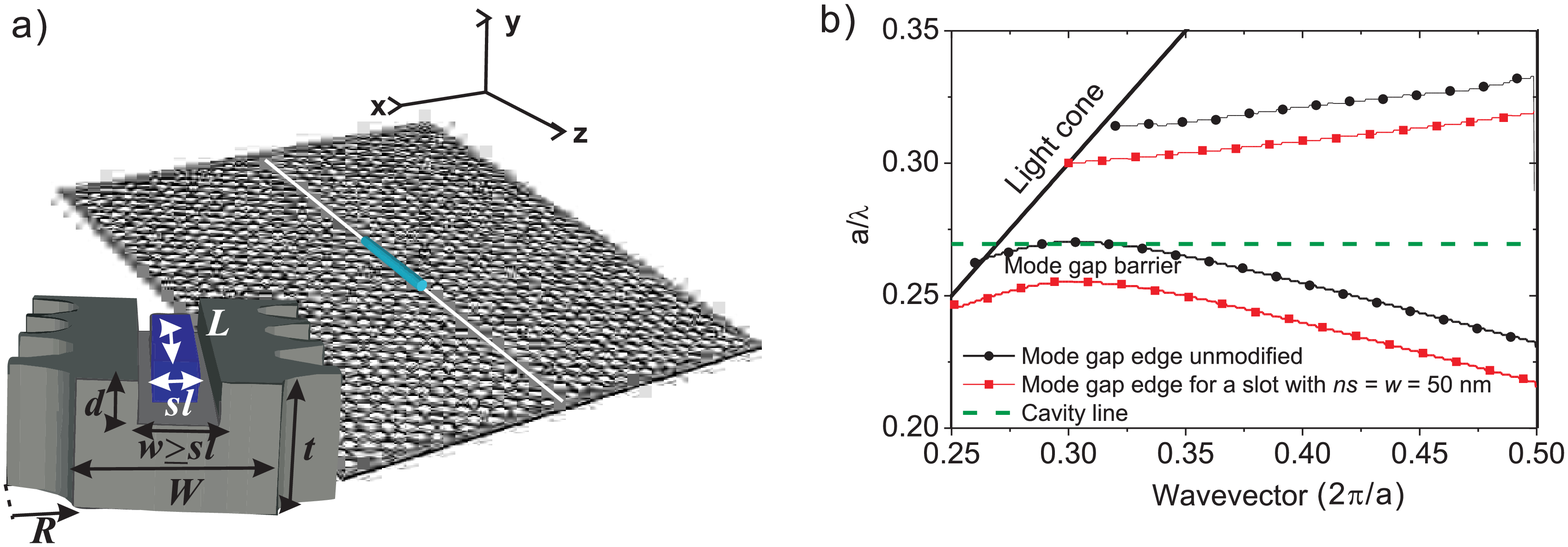}
\caption{(a) Schematic of a position-controlled ultrahigh Q-cavity using a single NW inside an air slot in a line defect of 2D photonic crystal. The inset shows the geometry of the sample. The white and the black letters indicate the parameters of the NW and the photonic crystal, respectively; a line defect width ($W$) of 0.98$a\sqrt{3}$, a hole radius ($R$) of 130 nm, a NW cross section side length ($sl$) of 50 nm, an air slot depth ($d$) between 25 and 75 nm, a slot width ($w$) between 50 and 200 nm, and a NW length ($L$) between 3 and {5}$\mu$m. (b) Photonic bandstructure calculation for photonic crystals with a lattice constant ($a$) of 380 nm with only a line defect (black circles) and that with a NW of $L$ = {3}$\mu$m and $sl$ = 50 nm and an air-slot of $w$ = 50 nm and $d$ = 50 nm (red squares). Other parameters are the same as in (a). A cavity formation created by introducing the air slot is shown by the green dashed line.}
\label{Fig1}
\end{figure}

In this work, we consider NWs with a refractive index ($n_{NW}$) of 3.17 (i.e. InP NWs), while the substrate has a refractive index ($n_{subs}$) of 3.48 (i.e. silicon (Si)). However, this approach is suitable for different material systems using different parameters. We assume a square cross-section for our NWs with a side length ($sl$) of 50 nm and lengths ($L$) between 3 and 5 $\mu$m (see Fig. \ref{Fig1}a). These are conventional parameters for InP NWs grown using low-pressure metalorganic vapor phase epitaxy (MOVPE) with [001] direction on InP (001) substrate \cite{Krishna2004}. We choose a NW with a square cross-section to simplify the geometry analysis needed for our calculations. In Fig. \ref{Fig1}a shows a photonic crystal design based on a line-defect waveguide (consisting of a row of missing holes in the $\Gamma-K$ direction). The hole radius ($R$) and the thickness ($t$) are constants of 130 and 150 nm, respectively, while the lattice constant ($a$) is 380 nm. The width of the line defect (the distance between the nearest neighbor holes on either side of the missing-hole row) is 0.98$\times{a}\sqrt{3}$. For the air slot, we vary the parameters such as the slot depth ($d$ = 25-75 nm), width ($w$ = 50-200 nm), and the position of the NW to investigate the change in $Q$ and $V_{c}$ caused by errors in slot fabrication or the nanopositioning of the NW.

In our design, mode-gap confined photonic crystal nanocavities are realized by the NW while the air slot is introduced as a dielectric discontinuity to enhance the electric field from the original cavities. The physical mechanism is very similar to that reported in previous studies on entirely-drilled air slots \cite{LipsonPRL2005,Takayuki2008}, in which the local field enhancement at dielectric boundaries is considered. However, unlike the cavity with one-sided cladding \cite{Tanaka2006}, the termination of the air slot does not cause much $Q$ degradation in spite of the strong vertical asymmetry in a 2D photonic crystal slab. The radiation loss to the lower part of the slab is suppressed by the NW in the center of the air slot, which increases the effective refractive index.

In general, the electric and magnetic fields of the transverse electric (TE) mode in a photonic crystal cavity are mainly $x-$ and $y-,z-$polarized, respectively. First, we consider electric field enhancement for the simplest photonic crystal design, where the NW side length matches the air-slot width. For maximum enhancement, $w$ should be infinitesimally narrow while $d$ should be shallow enough. It also means that the NW should be as thin as possible. A large slot and NW may reduce the field strength at the slot boundaries. This explains the choice of a NW with a 50-nm long side NW in our design since this size is suitably small for the NW. In the real design, the NW can not fit with the air slot perfectly. Then, the enhancement condition will be different in this case. For this reason, the $w$ and $d$ values of the air slot in our design must be optimized with the $sl$ of the NW so that the electric field of the cavity can be enhanced optimally. We also note that the misalignment of the NW from the center of the slot may reduce the field strength.

The presence of the air slot in our design has certain advantages regarding the difficulty involved in the high precision nanopositioning of the NW in the line defect. Placing a single nanowire in the center of a simple line defect in the photonic crystal can create a cavity but with a moderate $Q \lesssim 10^{4}$. However, the $Q$ is extremely sensitive with respect to the position of the NW, especially the $x$-direction(see Fig. \ref{Fig1}a). The exact position of the NW in the center of the line defect requires more precision than that provided by current nanopositioning technology. Therefore, the air slot acts as a barrier for the spatial movement of the NW. With the air slot, we can also move the NW in the line defect along the $z$-direction with a small error in the $x$-direction.

We can also control the refractive index of the line defect waveguide through the depth of the air slot. Theoretically, ultrahigh-$Q$ nanocavity modes can be created by the ultrasmall local modification of the refractive index of a line defect waveguide in a 2D photonic crystal slab \cite{Notomi2008}. This was demonstrated by employing a photosensitive process for chalcogenide glass photonic crystals \cite{Lee2009} and scanning probe lithography for Si photonic crystals \cite{Yokoo2011}. Here we propose that removing small amounts of Si in the line defect through the air slot affects the confinement around the NW. This mechanism is explained by the schematic band diagram in Fig. \ref{Fig1}b, which was calculated by the plane-wave expansion method. For this band diagram calculation, we set $a$ = 380 nm, $L$ = 3 $\mu$m, $R$ = 130 nm, and $sl$ = $w$ = $d$ = 50 nm. Without the NW and the air slot, the line defect in the photonic crystal has a sharp mode gap edge in the band diagram (see black dotted lines in Fig. \ref{Fig1}b). If we lower the mode-gap edge of the central part through the air slot and the NW (red dotted lines), we create a confined mode surrounded by the mode gap in an unmodified line defect (green dashed lines). Such modification is equivalent to structural \cite{BSSong2005,Kuramochi2006} and refractive-index modulations \cite{Notomi2008,Hanic2007}.

To investigate the field distributions, we performed a 3D FDTD with a calculation area as large as 32.4 ($x$ axis), 7.9 ($y$ axis) and 40.3 ($z$ axis) lattice periods. The perfectly matched layer method with 8 layers was employed as absorbed boundary condition in all directions. We set a grid spacing of 25 nm so that one of more field components was contained within the 50-nm NW side length. For the details of the method, we first performed the simulation using a broadband excitation light source and obtained a cavity spectrum through Fourier-transformed field. We defined the resonance angular frequency $\omega$ and used this parameter to simulate the field and the energy decay of the cavity mode with a narrow band excitation. To determine $Q$ which represents the strength of the light confinement, we fitted the tail of the energy decay curve with a single exponential and we determined the cavity photon lifetime $\tau_{ph}$. We estimated the Q in a cavity as $Q = \tau_{ph}\cdot\omega$.

\section{FDTD simulation for NW in width-matched air slot}

\begin{figure}[htbp]
\centering\includegraphics[width=10cm]{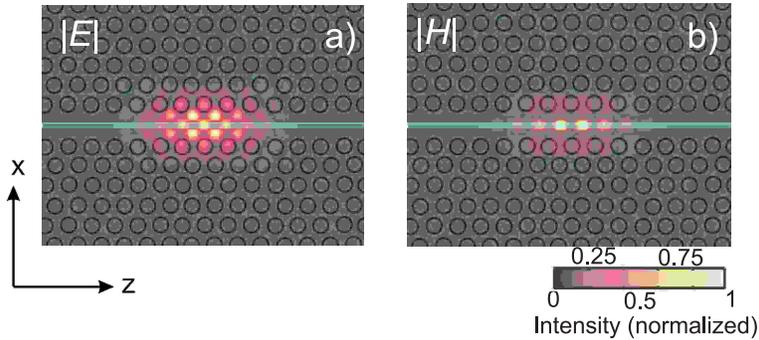}
\caption{(a) Calculated electric-field ($|E|$) and magnetic-field ($|H|$) distributions in the $x-z$ plane for $a$ = 380 nm, $L$ = 3 $\mu$m, $R$ = 130 nm, and $sl$ = $w$ = $d$ = 50 nm. The color scale is linear and blue filled and empty rectangles in the middle represent the NW and the air slot, respectively.}
\label{Fig2}
\end{figure}

First, we simulate the field distributions for a condition where the side length of the NW perfectly matches the air-slot width in the line defect of the photonic crystal $sl$ = $w$ = $d$ = 50 nm. Fig. \ref{Fig2}a, shows the electric-field ($|E|$) and magnetic-field ($|H|$) distributions in the x-z plane for the fundamental TE mode. The electric- and the magnetic-fields are mainly dominated by $x$- and $y$-polarizations, respectively, and both fields are confined in the 3-$\mu$m-length ~NW. For this cavity mode, the cavity wavelength is exist at the telecommunication wavelength (1410 nm) while $Q$ and $V_{c}$ are $4 \times 10^{5}$ and 0.07 $\mu$m$^{3}$, respectively. Although this design already exhibits high $Q$, an NW-width-matched air slot is difficult to realize experimentally, i.e. nanomanipulation of the NW requires a slot, that is larger than the NW.

\section{FDTD simulation for the NW with different lengths $L$ in the width-matched air slot}

\begin{figure}[htbp]
\centering\includegraphics[width=13cm]{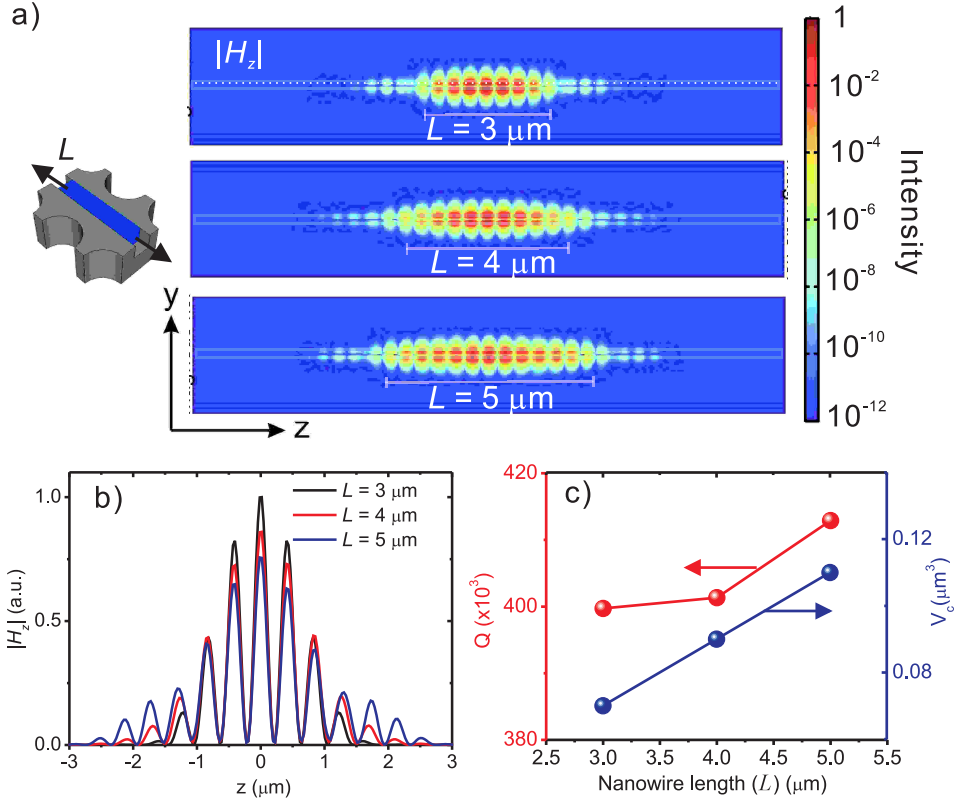}
\caption{(a) Calculated $z$-component of the magnetic-field ($|H_{z}|$) distributions in the $y-z$ plane for different $L$ values of the NW and other parameters; $a$ = 380 nm, $R$ = 130 nm, and $sl$ = $w$ = $d$ = 50 nm. The color scale is logarithmic and photonic crystal slabs are depicted by solid lines. The tick lines show the lengths of the NW and the relative positions of the NW in the slab. Each field is normalized by the total field energy in the whole calculation area. (b) The magnetic-field profile along the z direction (the maximum field) indicated by dotted lines in (a). (c) Summarized performance of $Q$ and $V_{c}$ for different $L$ values.}
\label{Fig3}
\end{figure}

It is important to simulate the sensitivity of the $L$ value of the NW with $Q$ and $V_{c}$. Principally, the NW length can be determined by controlling the growing condition but it is difficult to choose a precise length of 3-$\mu$m from many NWs with certain length distributions. Here we discuss the fidelity of $Q$ and $V_{c}$ with different $L$ values for our design from an analysis of the $z$-component of the magnetic-field ($H_{z}$) distributions in Fig. \ref{Fig3}a. When the NW becomes longer, the mode distributions become spread along the waveguide ($z$) direction. We found that the full widths at half maximum (FWHM) of the mode distributions were 1.46, 1.81, and 2.33 $\mu$m ~for 3-, 4-, and 5-$\mu$m ~NW, respectively. This shows that the spread of the mode distributions is proportional to $L$. From the profile along the waveguide depicted in Fig. \ref{Fig3}b, as $L$ increases, the intensity in the center decreases while that at the side increases. Quantitatively, $Q/V_{c}$ is relatively constant, which means that the enhancement is not sensitive to $L$ (see Fig. \ref{Fig3}c). The $Q$ itself has a small variation. This is most likely due to the small difference in the local index modulation caused by different $L$ in comparison with the large difference in the refractive index of NW and that of the photonic crystal \cite{Notomi2008}. We observed that the increase in $V_{c}$ is accompanied by the spreading of the mode distributions, which is similar with the case of the local index modulation in the waveguide of the photonic crystal \cite{Notomi2008}.

\section{FDTD simulation for NW with different slot depths $d$}

It is more difficult to fabricate an air slot with a partial slab depth ($d$) than an entirely-drilled air slot. An electron beam lithography or dry etching process may provide a different $d$ from that expected. Therefore, we include the $d$ variation ($d \pm 0.5 sl$) in our simulation for the investigation of $Q$ and $V_{c}$. In this calculation, other than $d$, we use the same parameters as in Fig. \ref{Fig2}a.

\begin{figure}[htbp]
\centering\includegraphics[width=13cm]{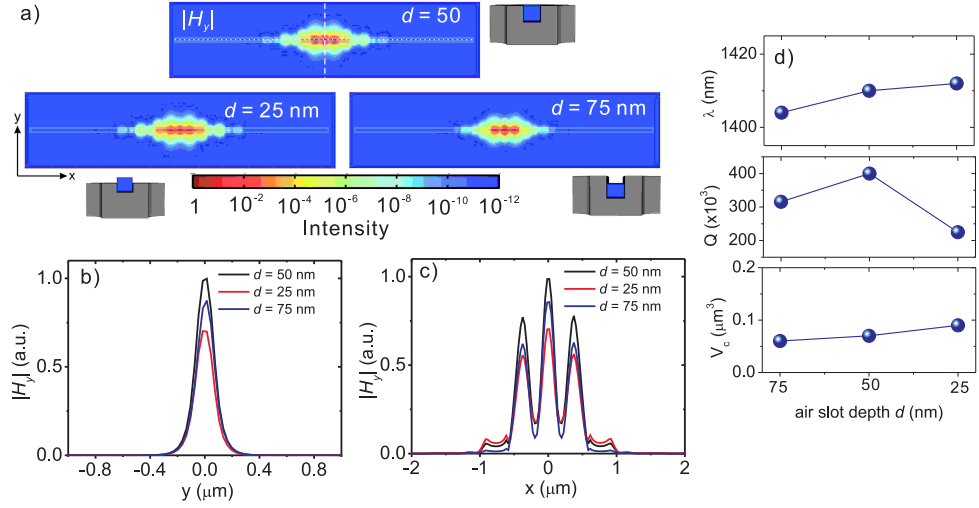}
\caption{(a) Calculated $y$-component of the magnetic-field ($|H_{y}|$) distributions in the $x-y$ plane for different $d$ values of the air slot and other parameters; $a$ = 380 nm, $R$ = 130 nm, $L$ = 3 $\mu$m, and $sl$ = $w$ = 50 nm. The color scale is logarithmic and photonic crystal slabs are depicted by solid lines. Each field is normalized by the total field energy in the whole calculation area. The magnetic-field profiles along (b) the $y$-direction and (c) the $x$-direction correspond to the vertical dashed and horizontal dotted lines, respectively, indicated in (a). (d) $d$ dependences of cavity peak $\lambda$, $Q$, and $V_{c}$.}
\label{Fig4}
\end{figure}

The magnetic-field distributions ($H_{y}$) in the $x-y$ plane are shown in \ref{Fig4}a. From these results, we observe that the field profiles along the $y$-direction are relatively similar while the profiles along the $x$-direction change with the position of the NW (see Fig. \ref{Fig4}b and c for $y$- and $x$-directions, respectively).  The mode distributions have the longest spread in the $x$-direction with the shallowest air slot. The profile becomes shorter for a deeper air slot. With both shallow and deep air slots, the intensity in the center decreases. The cavity peak, $Q$ and $V_{c}$ are summarized in Fig. \ref{Fig4}d. We notice that the cavity peak $\lambda$ exhibited an 8-nm-red shift with a change in $d$. Therefore, this design is interesting in terms of detuning the wavelength using an air slot whose depth varies along the $z$-direction. $V_{c}$ is relatively constant for different $d$ values. The highest $Q$ is found when the depth is the same as the NW side length $d = sl$. The maximum decrease in $Q$ is only by a factor 2 for $d = 1.5 sl$, which is still acceptable for fabrication with electron beam lithography \cite{Notomi2010} and scanning probe lithography \cite{Yokoo2011}.

\section{FDTD simulation for NW with different slot widths $w$}

\begin{figure}[htbp]
\centering\includegraphics[width=13cm]{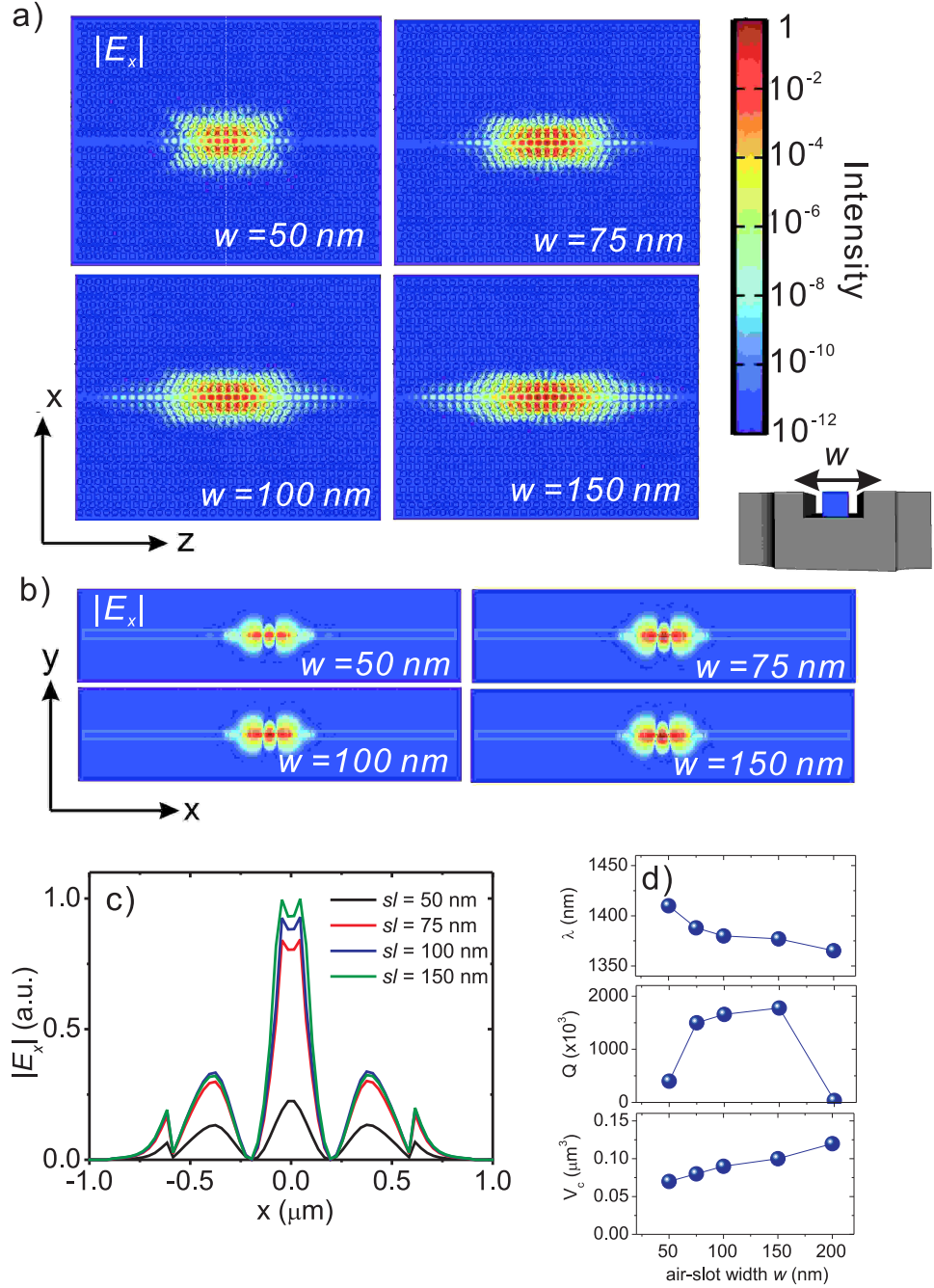}
\caption{(a) Calculated $x$-component of the electric-field ($|E_{x}|$) distributions in (a) the x-z and (b) x-y planes for different $w$ values of the air slot and other parameters; $a$ = 380 nm, $R$ = 130 nm, $L$ = 3 $\mu$m, and $sl$ = $d$ = 50 nm. The color scale is logarithmic and photonic crystal structures are depicted by solid lines. Each field is normalized by the total field energy in the whole calculation area. (c) The electric-field profile along the $x$-direction corresponds to the vertical dotted lines in (a). (d) Cavity peak $\lambda$, $Q$, and $V_{c}$ as a function of $w$.}
\label{Fig5}
\end{figure}

In our design, our aim is to set the optimum width of the air slot so that we can easily manipulate the NW in the center of the air slot without any loss of $Q$. Fig. \ref{Fig5}a and b show the simulated electric field $E_{x}$ for different $w$ values. The mode distibutions are very sensitive to $w$, especially from $w = 50$ nm to $w = 75$ nm. As $w$ becomes larger, we observe that the electric field is enhanced in the air slot while the mode distributions become narrow and spread in the waveguide direction. This is similar to the case with different $L$ values and an entirely drilled air slot described in Ref. \cite{Takayuki2008}. The transversal profile of $E_{x}$ along the $x$-direction is shown in Fig. \ref{Fig5}c. As with the entirely drilled air slot in Ref. \cite{Takayuki2008}, the electric-field maxima are positioned at the boundary of the NW because the enhanced field originates from polarization charges induced in the NW walls. This effect is more pronounced when the slot walls have a boundary with the air of $50$ nm$ < w \lesssim 150$ nm. The intensities at the slot walls and in the center increase along with the increase in $w$. Here we simulated the square cross-section NW, but the trend may not be so different for the other type of NW facet. Fig. \ref{Fig5}d shows a significant increase in $Q$ from the design where $50$ nm$ < w \lesssim 150$ nm. The largest $Q$ of 1.8$\times10^{6}$ for $w = 150$ nm constitutes approximately a 6-fold improvement compared with that of the NW with a 1D photonic crystal cavity proposed in Ref. \cite{Zhang2008}. For $w > 150$ nm, Q decreases as a result of the decrease in the refractive index, i.e. the condition for total internal reflection (TIR) becomes severe as the slot width increases \cite{Takayuki2008}. Additionally, we observe a blue shift of $\lambda$ and an increase in $V_{c}$ towards a large $w$ value. The $Q/V_{c}$ derived from Fig. \ref{Fig5}d is relatively constant for the design where $w$ ranges from 75 nm to 150 nm. This means that we can choose the slot width within this range. However, another factor for decreasing $Q$, which limits the selection of $w$, is discussed in the following section.

\section{FDTD simulation for the misalignments of the NW from the center of the slot}

\begin{figure}[htbp]
\centering\includegraphics[width=13cm]{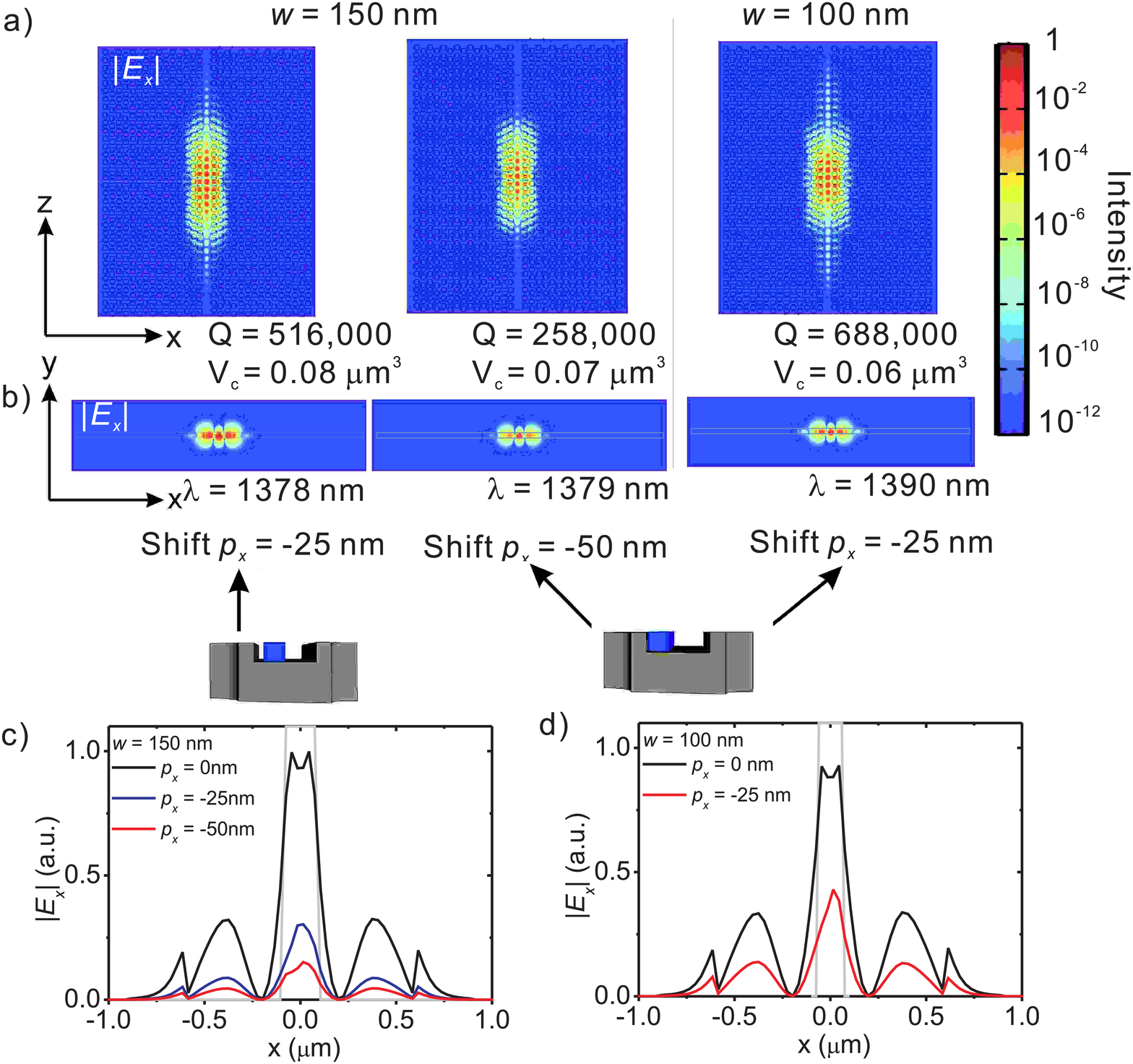}
\caption{(a) Calculated $x$-component of the electric field ($|E_{x}|$) distributions in (a) the x-z and (b) x-y planes for different $w$ values with a misalignment of the NW $p_{x}$ and other parameters; $a$ = 380 nm, $R$ = 130 nm, $L$ = 3 $\mu$m, and $sl$ = $d$ = 50 nm. The color scale is logarithmic and photonic crystal structures are depicted by solid lines. Each field is normalized by the total field energy in the whole calculation area. The electric-field profile along the $x$-direction corresponds to the horizontal dotted lines shown in (a) for (c) $w$ = 150 nm and (d) $w$ = 100 nm. Data for $p_{x}$ = 0 were obtained from Fig. \ref{Fig5}c and the grey lines show the air slot dimension.}
\label{Fig6}
\end{figure}

We have already shown that a large $Q$ can be obtained if the NW is combined with a large air slot. A large air slot has advantages in that it is easier to put NW into it. However, $Q$ depends on the position of the NW ($z$-direction). This means that a large air slot, despite its high $Q$, may experience a large decrease in $Q$ if the NW is not in the center of the air slot. Fig. \ref{Fig6}a and b show the calculated electric field $E_{x}$ for the NW inside 150- and 100-nm slots in the $x-z$ and $x-y$ planes, respectively. The spatial misalignment is indicated by $p_{x}$. We observe that the mode distributions depend strongly on the position of the NW and they become broader when the NW is closer to the slot wall. A narrow slot width ($w = 100$ nm) provides stronger light confinement than a wider slot ($w = 150$ nm) for the same NW spatial misalignment ($p_{x} = -25$ nm). The electric-field profiles for different spatial misalignments of $w = 150$ nm and $w = 100$ nm are shown in Fig. \ref{Fig6}c and d, respectively. The maximum of the field shifts to the opposite direction of the NW misalignment and the field strengths exhibit monotonic fade out for larger NW misalignments from the center of the slot. The cavity wavelength ($\lambda$) undergoes a red shift showing the potential for detuning using the NW position. When we compare the $Q$ of the maximum misalignment (the NW attached to one side of the slot walls $|p_{x}| = (w-sl)/2$) with that of the NW in the slot center, we find that the $Q$ exhibits an approximately 7-fold decrease for $w = 150$ nm while that of $w = 100$ nm exhibits a loss factor of 2.4. From Fig. \ref{Fig5}d, we found that the $Q/V_{c}$ values for the designs with $w$ between 75 and 150 nm are almost the same, while the results for the misalignment of the NW in $w = 100$ nm and $w = 150$ nm shows that a narrow air slot is better for maintaining a high-$Q$ in our design. Although we do not show the results for the misalignment of the NW for $w$ values between 75 and 100 nm, we may conclude that the optimum air-slot width for $Q/V_{c}$ can be found in this range.

\section{$k$-space field distribution of the cavity mode field}

\begin{figure}[htbp]
\centering\includegraphics[width=13cm]{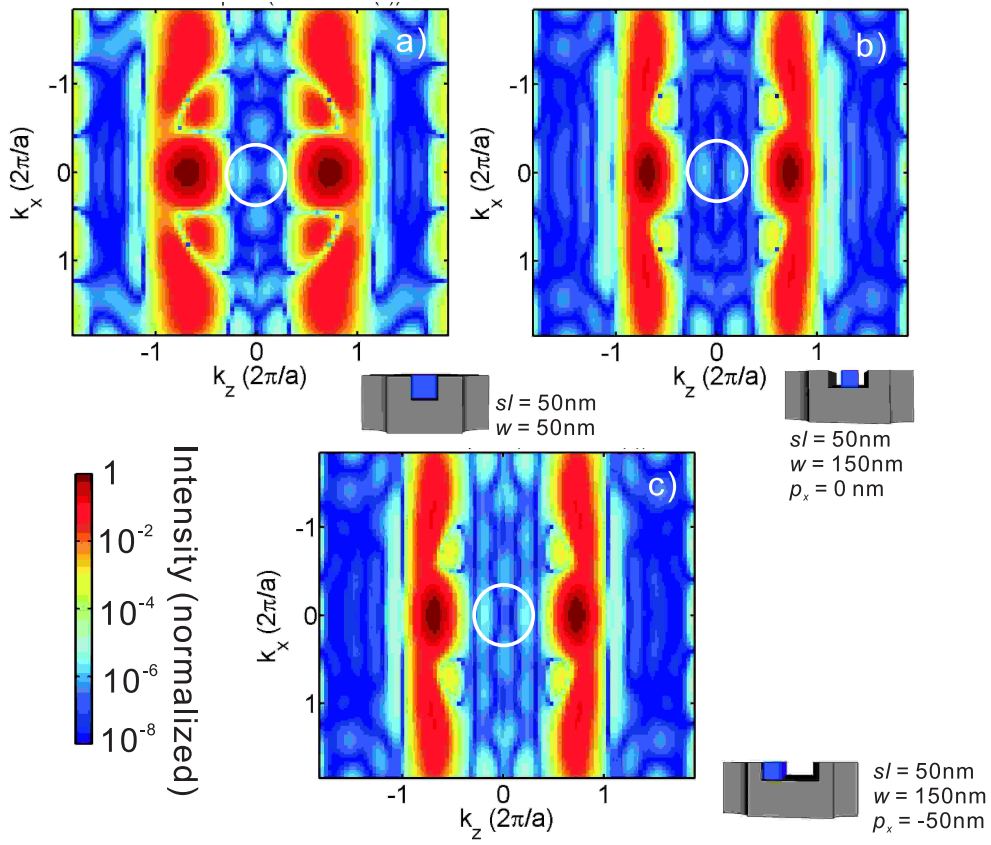}
\caption{Spatial Fourier transformation spectra obtained from the $x$-component of the electric field ($E_{x}$) in the x-z plane for (a) a NW in a width-matched slot ($sl = w = 50$ nm), (b) a NW in the center of a wide slot ($w = 150$ nm), and (c) a NW attached to one side of the slot walls ($w = 150$ nm, $p_{x} = -50$ nm). The leaky regions are shown by solid white lines. Other parameters used in the calculation are $a$ = 380 nm, $R$ = 130 nm, $L$ = 3 $\mu$m, and $d$ = 50 nm.}
\label{Fig7}
\end{figure}

Here we investigate the origin of high-$Q$ with a large air slot and the decrease in $Q$ with NW position misalignment through the $x-$ and $z-$directional spatial Fourier transformation (FT) of the cavity field. Fig. \ref{Fig7} shows the FT spectra of the field profile in the middle of a photonic crystal slab for (a) $sl = w = 50$ nm, (b) $sl = 50$ nm, $w = 150$ nm, $p_{x} = 0$ nm, and (c) $sl = 50$ nm, $w = 150$ nm, $p_{x} = -50$ nm. The light cones of air are represented by white circles with radii of 2$\pi/\lambda$. Inside these circles, the TIR condition is broken for decomposed plane momentum components ($k_{x},k_{z}$). As the slot width $w$ increases, $V_{c}$ increases and the number of momentum components inside the light cone decreases (see Fig. \ref{Fig7}a, b, and Fig. \ref{Fig5}d). This condition results in the 4.5-fold improvement in $Q$. On the other hand, the NW misalignment causes an increase in the number of momentum components inside the light cone (see Fig. \ref{Fig7}c). In this case, a 7-fold deterioration of $Q$ is observed (see Fig. \ref{Fig5}d and Fig. \ref{Fig6}a).

\section{Light-matter interaction of the embedded emitter NW}

Here we focus on the light-matter interaction of the embedded emitter NW (QD NW) inside the slot of the line defect of the photonic crystals. There have been many studies of heterostructure QD NW, mostly concerning InAs/GaAs and InAs/InAsP QD NWs \cite{Dorenbos2010,Claudon2010,Dalacu2011,Bleuse2011,Reimer2012,Bulgarini2012}. Of these studies, two reports show that a photonic NW can control the spontaneous emission \cite{Bleuse2011,Bulgarini2012}. They found that the spontaneous emission was greatly inhibited (between 12- and 16-fold) for a small-facet NW ($sl <$ 100 nm), from which they determined that the quantum efficiency of the QD NW was between 92 and 97 $\%$ \cite{Bleuse2011,Bulgarini2012}. In terms of the very efficient funnelling of the spontaneous emission into the guided mode, this photonic NW is generally to be an efficient single photon source. However, the inhibition of the spontaneous emission can be a drawback when using this small-facet photonic NW as a single photon source. Therefore, by using a photonic crystal cavity, we can expect the spontaneous emission rate of the small-facet NW to be enhanced.

For the following discussion, we will use the parameters of an InAs QD in InP NW since InAs/InP is already known to have potential as a single photon source in the telecom band \cite{Birowosuto2012}. To realize optimal coupling, we define the position of the QD as being the center of the NW while the transition dipole moment should be aligned with the electric field dipole. For a NW located in the center of the air slot, the center of the NW is also the location of the electric field maximum.

As first noted by Purcell \cite{Purcell1946}, the radiative emission rate of an emitter coupled with the resonant cavity mode can be derived directly from Fermi`s golden rule:
\begin{equation}
\Gamma = \frac{2\pi}{\hbar^{2}}\int_{-\infty}^{\infty}\left<|\vec{p}_{a}\cdot\xi\vec{E}(\vec{r}_{e})|^{2}\right>\rho_{c}(\omega)\rho_{e}(\omega)d\omega
\label{Fermi}
\end{equation}
where $\rho_{c}(\omega)$ is the density of photon modes in the cavity, $\rho_{e}(\omega)$ is the mode density for the dipole transition, $\vec{p}_{a}$ is the atomic dipole moment, and $\vec{E}(\vec{r}_{e})$ is the electric field at the location of the emitter normalized by a factor $\xi^{2} \equiv \frac{\hbar\omega}{2}\frac{4\pi}{\int\epsilon(\vec{r})\vec{E}^{2}(\vec{r})d^{3}r}$ so that it gives to the zero point energy.

The spontaneous emission rate can be enhanced in two ways. The first is to increase the cavity mode density $\rho_{c}(\omega)$, which is better known as $Q$. The other involves increasing the normalized electric field at the emitter position 
$(\xi\vec{E}(\vec{r}_{e}))$, which is proportional to the decrease in $V_{c}$. Therefore, the common figure of merit for the resonant cavities is the ratio $Q/V_{c}$, which can be seen from the Purcell factor $F_{p}$. In our case, we regard $F_{p}$ as the ratio of spontaneous emission rate of QD NW in the cavity $\Gamma_{QD-NW}^{cav}$ compared with that of the QD in the bulk $\Gamma_{QD}$;
\begin{eqnarray}
F_{p} & = & F_{p}^{cav-NW} \times \frac{\Gamma_{QD-NW}(sl/\lambda_{em})}{\Gamma_{QD}} = \frac{\Gamma_{QD-NW}^{cav}}{\Gamma_{QD}}\label{Pf1}\\
& = &  \frac{3Q\lambda^{3}}{4\pi^{2}{n_{NW}}^{3}}\frac{\epsilon(\vec{r}_{0})|\vec{E}(\vec{r}_{0})|^{2}}{\int{\epsilon(\vec{r})|\vec{E}(\vec{r})|^{2}d^{3}r}}\label{Pf2}\\
& = & \frac{3Q\lambda^{3}}{4\pi^{2}{n_{NW}}^{3}V_{c}} = \frac{6Q}{\pi^{2}\tilde{V}_{c}}\label{Pf3}
\end{eqnarray}
where $\vec{r}_{0}$ is the center of the photonic crystal slab, which is the peak field and $\lambda$ is the cavity wavelength. $F_{p}^{cav-NW}$ is the ratio of the spontaneous emission rate of a QD NW in the cavity compared with that for a QD NW without a cavity $\Gamma_{QD-NW}$, which depends on the facet size normalized to the emission wavelength of the NW ($sl/\lambda_{em}$). In a small-facet NW ($sl/\lambda_{em} < 0.16$), the emission rate is inhibited (${\Gamma_{QD-NW}}/{\Gamma_{QD}} < 1$) while in a large-facet NW ($sl/\lambda_{em} > 0.16$), the emission rate is slightly enhanced (${\Gamma_{QD-NW}}/{\Gamma_{QD}} \gtrsim 1$) \cite{Bleuse2011,Bulgarini2012}. We also define the dimensionless mode volume $\tilde{V}_{c}$ as
\begin{equation}
\tilde{V}_{c} = V_{c}\left(\frac{2n_{NW}}{\lambda}\right)^{3} 
\label{modevol}
\end{equation}

\begin{figure}[htbp]
\centering\includegraphics[width=13cm]{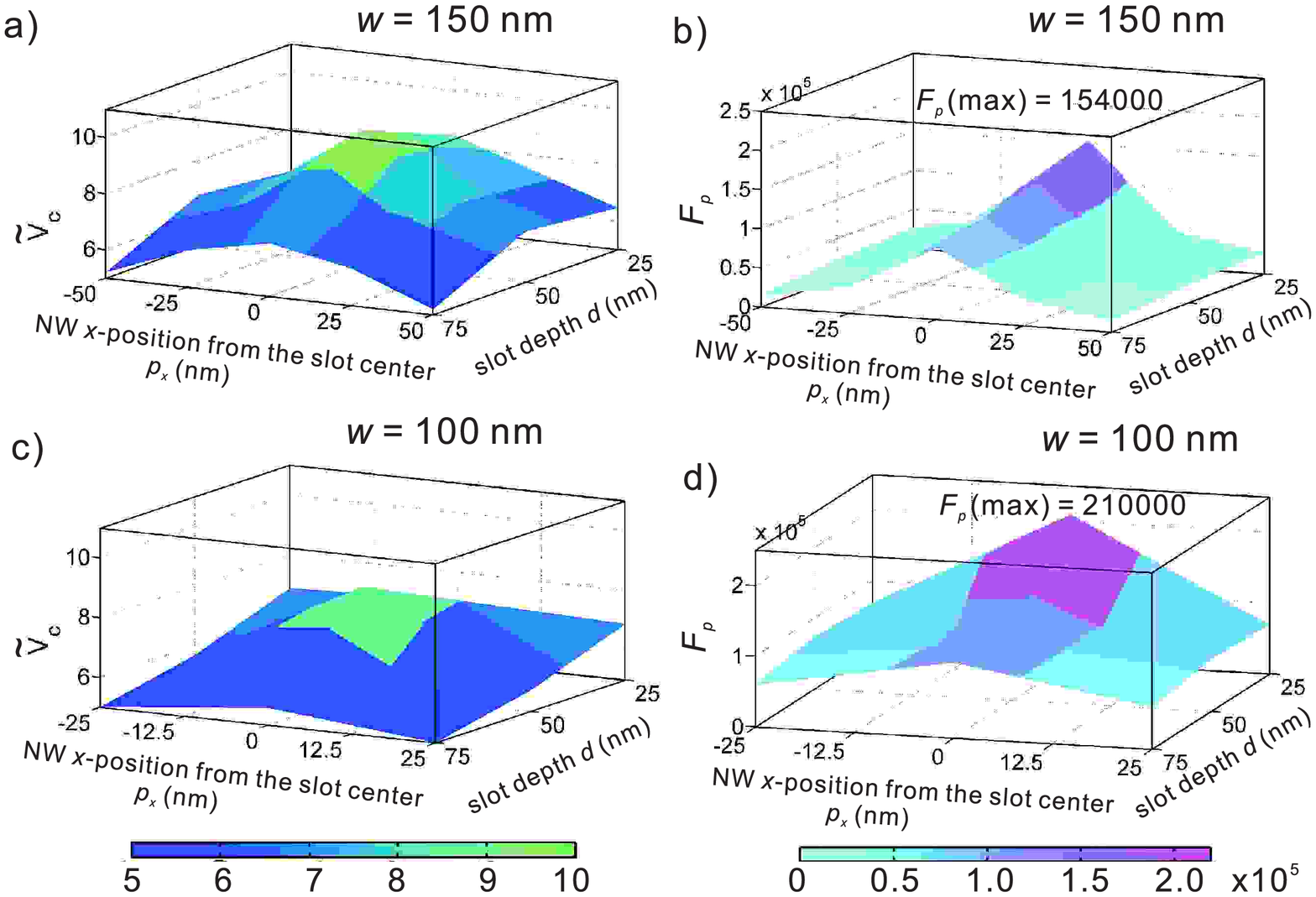}
\caption{(a) Dimensionless mode volume $\tilde{V_{c}}$ and (b) Purcell factor $F_{p}$ for a slot width $w$ of 150 nm calculated for different NW positions $p_{x}$ and slot depths $d$. (c) and (d) are $\tilde{V_{c}}$ and $F_{p}$ for $w = 100$ nm, respectively. Other parameters used in the calculation are $a$ = 380 nm, $R$ = 130 nm, $L$ = 3 $\mu$m, and $sl$ = 50 nm.}
\label{Fig8}
\end{figure}

Here we analyse the enhancement in the spontaneous emission rate caused by the increase in $Q$ for the large slot width as previously seen in Fig. \ref{Fig5}. Fig. \ref{Fig8} shows the dimensionless mode volume $\tilde{V_{c}}$ and the Purcell factor $F_{p}$ as functions of the position of the NW in the $x-$direction $p_{x}$ and the slot depth $d$ for slots with $w = 150$ and $100$ nm. For the slot with $w = 150$ nm, $\tilde{V_{c}}$ is slightly more sensitive than that in $w = 150$ nm (see Fig. \ref{Fig8}a and c). The minimum and the maximum $\tilde{V_{c}}$ values in the $w = 150$ nm slot are 5 and 10, respectively. Using a Q of between 3$\times$10$^{5}$ and 2.2$\times$10$^{6}$ and $\tilde{V_{c}}$, we obtain an $F_{p}$ of between 1.5$\times$10$^{4}$ and 2.1$\times$10$^{5}$ (see Fig. \ref{Fig8}b and d for $w = 150$ and $100$ nm, respectively). The position of the NW in the $x-$direction has a stronger influence on $F_{p}$ than the slot depth. For both large slots, the $F_{p}$ maxima are found for the NW in the center of the slot $p_{x} = 0$ nm, and the shallow depth slot $d = 25$ nm. The slot with $w = 100$ nm exhibits a more stable $F_{p}$ with $p_{x}$ than the slot with $w = 150$ nm. Here, we found the best $F_{p}$ of 2.1$\times$10$^{5}$ for $w$ = 100 nm, $p_{x}$ = 0 nm, and $d$ = 25 nm, but we still may obtain a larger $F_{p}$ for 75 nm $< w <$ 100 nm.

As we discussed in relation to Eq. \ref{Pf1}, $F_{p}$ in our calculation includes $F_{p}^{cav}$ and ${\Gamma_{QD-NW}}/{\Gamma_{QD}}$. This means that we can compare the emission rate enhancement for a QD NW before and after it is placed in the air-slot. Taking account $sl$ = 50 nm, $\lambda_{em}$ = 1400-1600 nm, and assuming high quantum efficiency of the QD NW ($> 90\%$), we can estimate that ${\Gamma_{QD-NW}}/{\Gamma_{QD}} < 0.0625$ \cite{Bleuse2011}. This estimation yields $F_{p}^{cav} >$ 3.4$\times$10$^{6}$ and such an enhancement will greatly improve the single photon rate of the small size NW. Thanks to the large $F_{p}$ and $F_{p}^{cav}$, the present design provides a very good platform for cavity QED with a QD. We can expect a greater enhancement of the spontaneous emission rate for a QD at telecommunication wavelengths than that reported in Ref. \cite{Birowosuto2012}.

The NW-photonic cavity may even enter the strong-coupling regime of light-matter interaction, where there is a coherent exchange of energy between the photon trapped in the cavity and the exciton trapped in the QD \cite{Khitrova2006,Gibbs2011}. This occurs when the cavity field $\kappa$, the QD exciton decay rate $\gamma$, the phonon dephasing rate $\gamma_{dp}$, and the cavity-QD frequency detuning $\delta$ are smaller than the exciton-photon coupling strength $g$;
\begin{equation}
\frac{\kappa+\gamma}{2} + \gamma_{dp} + i\delta < 2g
\label{strongcoup}
\end{equation}
where $g$ is simply written as
\begin{equation}
g = \frac{\Gamma_{QD}}{n_{NW}}\sqrt{\frac{3c\lambda^{2}}{(2\pi\Gamma_{QD})4V_{c}}}
\end{equation}
Assuming that the radiative lifetime of the exciton of QD is 2 ns \cite{Bulgarini2012,Birowosuto2012}, which is converted into a radiative decay rate  $\Gamma_{QD}$ of 0.5 GHz, we obtain a $g$ value of 170 GHz. The non-radiative recombination rate of the QD is about 42 MHz \cite{Bulgarini2012,Birowosuto2012}, which is much lower than $g$. The dephasing rate $\gamma_{dp}$ for the telecom-band QD is typically 290 MHz \cite{Hayase2006}, and thus the only limiting factor shown in Eq. \ref{strongcoup} is $\kappa$. In our design, the smallest $Q$ for the NW in the large slot is about 3$\times$10$^{5}$. This is equal to $\kappa = 648$ MHz, which is still much smaller than 170 GHz. Therefore, we note that our design is suitable for the strong coupling regime even when the NW is misplaced in the slot of the photonic crystal.

\section{Conclusion}

We describe a novel photonic crystal cavity design, which is suitable for a position-controlled cavity for on-demand photonic controlled devices in photonic integrated circuits. This is the first design to realize a position-controlled cavity with a high $Q$ of $\simeq 2 \times $10$^{6}$ and $\tilde{V}_{c} < 10$. This design also shows that $Q$ and $V_{c}$ are not sensitive to the fabrication of the slot (depth and width), the NW size, and an misalignment of the NW position. Our design is also useful for a cavity QED platform with an embedded QD NW. This system introduces the concept of the position-controlled cavity and emitter, which can be useful for coupling the emitter and the cavity. The NW position dependence of $Q$ and the cavity wavelength are interesting as regards controlling the interaction between the cavity and the emitter. In terms of application, our system could be useful for the position-controlled devices such as position-controlled single photon sources, nanolasers, light emitting devices, and coupled cavities, though we need to do further investigation on the $Q$ dependence with the spatial translation of the NW along the line defect. As regards fabrication, our approach requires only simple procedures involving the nanomanipulation of an NW inside a slot in a line defect of a photonic crystal.

\section{Acknowledgment}
We are grateful for stimulating discussions with G. Zhang, T. Tawara, A. Shinya, H. Sumikura, J. Kim, K. Nozaki, M. Ono, and H. Xu. Part of this work was supported by Core Research for Evolutional Science and Technology-Japan Science and Technology Agency (CREST-JST).

\begin{thebibliography}{40}
\bibitem{Vahala2003}
K.~J. Vahala, Nature \textbf{424}, 839--46
  (2003).

\bibitem{Notomi2010}
M.~Notomi, Rep. on Prog. in Phys. \textbf{73}, 096501 (2010).

\bibitem{BSSong2005}
B.~Song, S.~Noda, T.~Asano, and Y.~Akahane, {Nature Materials} \textbf{{4}},
  {207--210} ({2005}).

\bibitem{Kuramochi2006}
E.~Kuramochi, M.~Notomi, S.~Mitsugi, A.~Shinya, T.~Tanabe, and T.~Watanabe,
 Appl. Phys. Lett. \textbf{88}, 041112
  (2006).

\bibitem{Tanabe2007}
T.~Tanabe, M.~Notomi, E.~Kuramochi, A.~Shinya, and H.~Taniyama,
 Nature Phot. \textbf{1}, 49--52 (2007).

\bibitem{Shinya2008}
A.~Shinya, S.~Matsuo, Yosia, T.~Tanabe, E.~Kuramochi, T.~Sato, T.~Kakitsuka,
  and M.~Notomi, Opt. Express \textbf{16}, 19382--19387 (2008).

\bibitem{Nozaki2010}
K.~Nozaki, T.~Tanabe, A.~Shinya, S.~Matsuo, T.~Sato, H.~Taniyama, and
  M.~Notomi, Nature Phot. \textbf{4}, 477--483 (2010).

\bibitem{Matsuo2010}
S.~Matsuo, A.~Shinya, T.~Kakitsuka, K.~Nozaki, T.~Segawa, T.~Sato,
  Y.~Kawaguchi, and M.~Notomi, Nature Phot. \textbf{4}, 648--654 (2010).

\bibitem{Ellis2011}
B.~Ellis, M.~A. Mayer, G.~Shambat, T.~Sarmiento, J.~Harris, E.~E. Haller, and
  J.~Vuckovic, {Nature Phot.} \textbf{{5}}, {297--300}
  ({2011}).

\bibitem{Nozaki2012}
K.~Nozaki, A.~Shinya, S.~Matsuo, Y.~Suzaki, T.~Segawa, T.~Sato, Y.~Kawaguchi,
  R.~Takahashi, and M.~Notomi, Nature Phot. \textbf{6}, 248--252 (2012).

\bibitem{Notomi2004}
M.~Notomi, A.~Shinya, S.~Mitsugi, E.~Kuramochi, and H.~Ryu,
  Opt. Express \textbf{12}, 1551--1561 (2004).

\bibitem{Yokoo2011}
A.~Yokoo, T.~Tanabe, E.~Kuramochi, and M.~Notomi, Nano Letters \textbf{11}, 3634--3642
  (2011).

\bibitem{Intonti2006}
F.~Intonti, S.~Vignolini, V.~T\"{u}rck, M.~Colocci, P.~Bettotti, L.~Pavesi,
  S.~L. Schweizer, R.~Wehrspohn, and D.~Wiersma, Appl. Phys. Lett. \textbf{89}, 211117 (2006).

\bibitem{Intonti2009}
F.~Intonti, S.~Vignolini, F.~Riboli, M.~Zani, D.~S. Wiersma, L.~Balet, L.~H.
  Li, M.~Francardi, A.~Gerardino, A.~Fiore, and M.~Gurioli, Appl. Phys. Lett. \textbf{95}, 173112 (2009).

\bibitem{Barrelet2006}
C.~J. Barrelet, J.~Bao, M.~Loncar, H.-G. Park, F.~Capasso, and C.~M. Lieber,
   Nano Letters \textbf{6}, 11--15 (2006).

\bibitem{Zhang2008}
Y.~Zhang and M.~Loncar, Opt. Express \textbf{16}, 17400--17409
  (2008).

\bibitem{Grillet2007}
C.~Grillet, C.~Monat, C.~L. Smith, B.~J. Eggleton, D.~J. Moss,
  S.~Fr\'{e}d\'{e}rick, D.~Dalacu, P.~J. Poole, J.~Lapointe, G.~Aers, and R.~L.
  Williams, Opt. Express \textbf{15}, 1267--1276 (2007).

\bibitem{Park2008}
H.-G. Park, C.~J. Barrelet, Y.~Wu, B.~Tian, F.~Qian, and C.~M. Lieber,
 {Nature Phot.} \textbf{{2}}, {622--626} ({2008}).

\bibitem{Dorenbos2010}
S.~N. Dorenbos, H.~Sasakura, M.~P. van Kouwen, N.~Akopian, S.~Adachi,
  N.~Namekata, M.~Jo, J.~Motohisa, Y.~Kobayashi, K.~Tomioka, T.~Fukui,
  S.~Inoue, H.~Kumano, C.~M. Natarajan, R.~H. Hadfield, T.~Zijlstra, T.~M.
  Klapwijk, V.~Zwiller, and I.~Suemune, Appl. Phys. Lett. \textbf{97}, 171106
  (2010).

\bibitem{Claudon2010}
J.~Claudon, J.~Bleuse, N.~S. Malik, M.~Bazin, N.~Gregersen, C.~Sauvan,
  P.~Lalanne, and J.-M. Gerard, Nature Phot. \textbf{4}, 174--177 (2010).

\bibitem{Dalacu2011}
D.~Dalacu, K.~Mnaymneh, X.~Wu, J.~Lapointe, G.~C. Aers, P.~J. Poole, and R.~L.
  Williams, Appl. Phys. Lett.  (2011).

\bibitem{Bleuse2011}
J.~Bleuse, J.~Claudon, M.~Creasey, N.~S. Malik, J.-M. G\'erard, I.~Maksymov,
  J.-P. Hugonin, and P.~Lalanne, Phys. Rev. Lett.
  \textbf{106}, 103601 (2011).

\bibitem{Reimer2012}
M.~E. Reimer, G.~Bulgarini, N.~Akopian, M.~Hocevar, M.~B. Bavinck, M.~A.
  Verheijen, E.~P. Bakkers, L.~P. Kouwenhoven, and V.~Zwiller, Nat. Commun.
  \textbf{3}, 737 (2012).

\bibitem{Bulgarini2012}
G.~Bulgarini, M.~E. Reimer, T.~Zehender, M.~Hocevar, E.~P. A.~M. Bakkers, L.~P.
  Kouwenhoven, and V.~Zwiller, Appl. Phys. Lett.
  \textbf{100}, 121106 (2012).

\bibitem{Babinec2010}
T.~M. Babinec, B.~J.~M. Hausmann, M.~Khan, Y.~Zhang, J.~R. Maze, P.~R. Hemmer,
  and M.~Loncar, {Nature
  Nanotech.} \textbf{{5}}, {195--199} ({2010}).

\bibitem{Knill2001}
E.~Knill, R.~Laflamme, and G.~J. Milburn, Nature \textbf{409}, 46--52 (2001).

\bibitem{Gisin2002}
N.~Gisin, G.~Ribordy, W.~Tittel, and H.~Zbinden, Rev. Mod. Phys. \textbf{74}, 145--195 (2002).

\bibitem{Strauf2010}
S.~Strauf, Nature Phot. \textbf{4}, 132--134 (2010).

\bibitem{Krishna2004}
U.~Krishnamachari, M.~Borgstrom, B.~J. Ohlsson, N.~Panev, L.~Samuelson,
  W.~Seifert, M.~W. Larsson, and L.~R. Wallenberg, Appl. Phys. Lett.
  \textbf{85}, 2077--2079 (2004).

\bibitem{LipsonPRL2005}
J.~T. Robinson, C.~Manolatou, L.~Chen, and M.~Lipson, Phys. Rev. Lett. \textbf{95},
  143901 (2005).

\bibitem{Takayuki2008}
T.~Yamamoto, M.~Notomi, H.~Taniyama, E.~Kuramochi, Y.~Yoshikawa, Y.~Torii, and
  T.~Kuga, Opt. Express
  \textbf{16}, 13809--13817 (2008).

\bibitem{Tanaka2006}
Y.~Tanaka, T.~Asano, R.~Hatsuta, and S.~Noda, Appl. Phys. Lett.  (2006).

\bibitem{Notomi2008}
M.~Notomi and H.~Taniyama, Opt. Express \textbf{16},
  18657--18666 (2008).

\bibitem{Lee2009}
M.~W. Lee, C.~Grillet, S.~Tomljenovic-Hanic, E.~C. M\"{a}gi, D.~J. Moss, B.~J.
  Eggleton, X.~Gai, S.~Madden, D.-Y. Choi, D.~A.~P. Bulla, and
  B.~Luther-Davies, Opt. Lett. \textbf{34}, 3671--3673
  (2009).

\bibitem{Hanic2007}
S.~Tomljenovic-Hanic, M.~J. Steel, C.~M. de~Sterke, and D.~J. Moss,
 Opt. Lett. \textbf{32}, 542--544 (2007).

\bibitem{Birowosuto2012}
M.~D. Birowosuto, H.~Sumikura, S.~Matsuo, H.~Taniyama, P.~J. van Veldhoven,
  R.~Noetzel, and M.~Notomi, {Sci. Rep.} \textbf{{2}}, {321} ({2012}).

\bibitem{Purcell1946}
E.~M. Purcell, Phys. Rev. \textbf{69}, 681 (1946).

\bibitem{Khitrova2006}
G.~Khitrova, H.~M. Gibbs, M.~Kira, S.~W. Koch, and A.~Scherer, Nature Phys. \textbf{2}, 81--90 (2006).

\bibitem{Gibbs2011}
H.~M. Gibbs, G.~Khitrova, and S.~W. Koch, Nature Phot. \textbf{5}, 273--273 (2011).

\bibitem{Hayase2006}
J.~Ishi-Hayase, K.~Akahane, N.~Yamamoto, M.~Sasaki, M.~Kujiraoka, and K.~Ema, Appl. Phys. Lett. \textbf{88}, 261907 (2006).
\end{thebibliography}
\end{document}